\documentclass[a4paper, UKenglish, cleveref, autoref, thm-restate]{lipics-v2021}
\hideLIPIcs  

\usepackage{bm}
\usepackage{dsfont}
\usepackage{xcolor}
\usepackage{stmaryrd}
\usepackage{amsthm}
\usepackage[tight-spacing=true]{siunitx}
\usepackage[linesnumbered,ruled,vlined]{algorithm2e}

\newcommand{\universe}{\ensuremath{\mathcal{U}}}
\newcommand{\family}{\ensuremath{\mathcal{F}}}
\newcommand{\stream}{\ensuremath{\mathcal{S}}}
\newcommand{\alg}{\ensuremath{\mathsf{MACH}}}
\newcommand{\algfull}{\ensuremath{\alg_{\lceil 2 \lambda \rceil}}}
\newcommand{\algopt}{\ensuremath{\alg_{\lfloor \varepsilon^2 \lambda / 3 \rfloor}}}
\newcommand{\algpair}{\ensuremath{\alg_2}}
\newcommand{\algfs}{\ensuremath{\alg_\text{fs}}}
\newcommand{\tp}{\ensuremath{\mathsf{TP}}}

\newtheorem*{corollary5}{Corollary 5}
\newtheorem*{lemma6}{Lemma 6}
\newtheorem*{proposition7}{Proposition 7}

\let\oldnl\nl
\newcommand{\nonl}{\renewcommand{\nl}{\let\nl\oldnl}}

\newlength\mylen

\usepackage{cite} 
\title{Maximum Coverage in Sublinear Space, Faster} 

\author{Stephen Jaud}{School of Computing and Information Systems, The University of Melbourne, Australia}{s.jaud@unimelb.edu.au}{https://orcid.org/0000-0002-8628-4512}{} 
\author{Anthony Wirth}{School of Computing and Information Systems, The University of Melbourne, Australia}{awirth@unimelb.edu.au}{https://orcid.org/0000-0003-3746-6704}{}
\author{Farhana Choudhury}{School of Computing and Information Systems, The University of Melbourne, Australia}{farhana.choudhury@unimelb.edu.au}{https://orcid.org/0000-0001-6529-4220}{}

\authorrunning{S.~Jaud, A.~Wirth, F.~Choudhury}

\Copyright{Stephen Jaud, Anthony Wirth, Farhana Choudhury}

\ccsdesc[100]{Theory of computation~Streaming, sublinear and near linear time algorithms}

\keywords{streaming algorithms, subsampling, maximum set cover, $k$-wise independent hash functions}

\funding{This research was supported by the Australian Government through the Australian Research Council's Discovery Projects funding scheme (project DP190102078). Farhana Choudhury is a recipient of the ECR22 grant from The University of Melbourne.}

\acknowledgements{Rowan Warneke, for reading and advising on an earlier version.}

\nolinenumbers

\begin{document}

\maketitle

\begin{abstract}

Given a collection of $m$ sets from a universe $\mathcal{U}$, the \textit{Maximum Set Coverage} problem consists of finding $k$ sets whose union has largest cardinality. This problem is NP-Hard, but the solution can be approximated by a polynomial time algorithm up to a factor $1-1/e$. However, this algorithm does not scale well with the input size.

In a streaming context, practical high-quality solutions are found, but with space complexity that scales linearly with respect to the size of the universe $n=|\mathcal{U}|$. However, one randomized streaming algorithm has been shown to produce a $1-1/e-\varepsilon$ approximation of the optimal solution with a space complexity that scales only poly-logarithmically with respect to~$m$ and~$n$. In order to achieve such a low space complexity, the authors used two techniques in their multi-pass approach:
\begin{itemize}
    \item \textit{Subsampling}, consists of only solving the problem on a subspace of the universe. It is implemented using $\gamma$-independent hash functions.
    \item $F_0$-\textit{sketching}, allows to determine with great accuracy the number of distinct elements in a set using less space than the set itself. 
\end{itemize}

This article focuses on the sublinear-space algorithm and highlights the time cost of these two techniques, especially subsampling. We present optimizations that significantly reduce the time complexity of the algorithm. Firstly, we give some optimizations that do not alter the space complexity, number of passes and approximation quality of the original algorithm. In particular, we reanalyze the error bounds to show that the original independence factor of $\Omega(\varepsilon^{-2} k \log m)$ can be fine-tuned to $\Omega(k \log m)$.
Secondly we show that $F_0$-\textit{sketching} can be replaced by a much more simple mechanism.

Although the theoretical guarantees are weaker, suggesting the approximation quality would suffer, for large streams, our algorithms perform well in practice. Finally, our experimental results show that even a pairwise-independent hash-function sampler does not produce worse solution than the original algorithm, while running significantly faster by several orders of magnitude.
\end{abstract}

\clearpage

\setcounter{page}{1}

\section{Introduction}
\label{sec:introduction}
\textit{Maximum Coverage}, also known as \textit{Maximum-$k$-Coverage} is a classic problem in computer science. Unless $\text{P}=\text{NP}$, the decision version is unsolvable in polynomial time. 
The input is a \textbf{family} of~$m$ sets, $\family$, each a subset of universe~$\universe$, comprising~$n$ elements, and a positive \textbf{integer},~$k$.
The task is to find a subfamily of~$k$ sets in~$\family$ whose union has largest cardinality.
The best-known polynomial-time approximation algorithm for \textit{Max Coverage} and the ``dual'' \textit{Set Cover} problem\footnote{In Set Cover, the aim is to return a subfamily of \emph{minimum cardinality} whose union is~$\universe$.}, is a greedy approach. For Max Coverage, the greedy algorithm has been shown to return a solution whose coverage is at least a $1-1/e$ approximation of the optimal solution. This is known to be asymptotically optimal~\cite{Feige1998}.

In practice, the greedy algorithm is much more effective than its theoretical guarantee would suggest, and typically produces a near-optimal solution on realistic inputs~\cite{Grossman1997}.
However, the greedy algorithm does not scale well with the size of the input.
In the last~15 years, there has been increasing interest in efficient implementation of greedy and greedy-like approaches for Set Cover and Maximum Coverage~\cite{Saha2009, Cormode2010, chakrabarti2016incidence}.
In the streaming setting, there have been several innovative algorithms, as detailed below in Table~\ref{tab:streaming-alg}. 
We focus in this paper on engineering the only sublinear-space Set Streaming algorithm~\cite{McGregor2018} so that it runs much faster, and sacrifices no space.

In the Set Streaming model~\cite{Saha2009}, the input stream, \stream, comprises a sequence of the sets in~\family, i.e., $\stream = S_1,S_2,\ldots,S_m$.
Each set~$S_i$ in~\stream\ appears in full before the next set,~$S_{i+1}$, appears.
The design of a streaming algorithm trades off memory, throughput, query/solution time, and solution quality.
Let~$I$ denote the indexes of the sets in the solution (so far).
The coverage of (the sets in)~$I$ is~$C = \cup_{i \in I} S_i$.
Given~$I$, and hence~$C$, the \emph{contribution} of each set~$S_j$, for every~$j \notin I$ is~$S_j \setminus C$.
In the greedy algorithm, we add a set to the solution whenever it has largest contribution, breaking ties arbitrarily. 
Additionally, another well-studied variant of this streaming model is \textit{random set arrivals}, a reasonable assumption for many applications, and it makes the problem \textit{easier}. Many results regarding trade-off between space complexity and approximation factor improve upon the classic \textit{set arrival} setting~\cite{NorouziFard2018, Agrawal2018}.
Another common model for Maximum Coverage, although not discussed in this paper, is the Edge-arrival Streaming model.
Here the stream consists of pairs $(i, x) \in [m] \times \universe$ to indicates that $x \in S_i$. In this more general context, Indyk and Vakilian \cite{Indyk2019} showed a space lower bound~$\Omega(m a^2)$ and upper bound $\Tilde{\mathcal{O}}(m a^2)$ for an arbitrary factor $a$-approximation factor in single pass.

\subsection{Sublinear Space}
Several of the greedy-like approaches for Set Cover in the Set Streaming model assume~$\Omega(n)$ memory is available~\cite{Cormode2010, lim2014lazy,chakrabarti2016incidence}: at least one bit per item, to record the coverage, and thus determine a set's contribution.
Unlike Maximum Coverage, in Set Cover, we expect that the subfamily of sets returned, indexed by~$I$, covers all of~$\universe$, so~$n=|\universe|$ bits seem necessary. In contrast, for Max Coverage, the minimum space requirement seems depend on~$m$. For example, it is known that every one-pass $(1/2+\varepsilon)$-approximation algorithm must work in~$\Omega(\varepsilon m /k^3)$ space~\cite{Feldman2020}. Also,~$\Omega(m)$ space is necessary to achieve better approximation factors than~$1-1/e$ \cite{McGregor2018}. Regarding $(1-\varepsilon)$-approximation algorithms, Assadi~\cite{Assidi2017} showed that $\Tilde{\Omega}(m/\varepsilon^2)$ space is required. It should be noted that all these lower bounds are tight and several one-pass $\Tilde{\mathcal{O}}(\varepsilon^{d} m)$-space algorithms do exist~\cite{bateni2017almost, McGregor2018}.

In this context, one algorithm for Maximum Coverage stands out.
McGregor and Vu~\cite{McGregor2018} introduced a family of streaming algorithms for Max Coverage.
They describe, in \S2.2 of their paper, an approximation algorithm that in $\mathcal{O}(\varepsilon^{-1})$ passes and in~$\Tilde{\mathcal{O}}(\varepsilon^{-2}k)$ space returns a $(1-1/e-\varepsilon)$-approximate solution\footnote{In this paper, the~$\Tilde{\mathcal{O}}(\cdot)$ notation hides polylogarithmic factors in~$m$ and~$n$.}.
This is the only reasonable approximation algorithm for Max Coverage that runs in~$o(\min\{m,n\})$ space.
For convenience\footnote{\alg\ represents ``\textsf{M}aximum {\sf A}ndrew {\sf C}overage {\sf H}oa'': the~$*$ represents their parameter choices, which we generalize in this paper.}, we name this algorithm~$\alg_*$.
Like some of the first streaming/external-memory algorithms for Set Cover, $\alg_*$ takes multiple passes, achieving a near-greedy approach via a sequence of decreasing thresholds for the contribution of a set: further details of thresholding are in \S\ref{sec:thresholding}.
And to save space, $\alg_*$ has a randomized subsampling component achieved with multi-way independent hash functions.
These hash functions are slow to evaluate, and it is this component that we accelerate.

\subsection{Motivation}
In terms of approximation quality and space complexity,   $\alg_*$ is the favored approach for Maximum Coverage.
The space complexity of $\Tilde{\mathcal{O}}(\varepsilon^{-2} k)$ is only a little more than the space required to
store\footnote{An approach that avoids storing at least one bit per index in~$I$, as working space, is in principle possible. For example,~$I$ could be a size-$k$ subset of~$\{1,\ldots,m\}$ chosen uniformly at random; this is not an effective solution, but a valid one, generated in~$\Tilde{\mathcal{O}}(1)$ working space.} solution~$I$: $\Tilde{\Omega}(k)$.
However, in   $\alg_*$, McGregor and Vu~\cite{McGregor2018} invoke a $\gamma$-independent hash function, where $\gamma = \lceil 2c \varepsilon^{-2} k \log m \rceil$, with~$c$ a constant to be discussed in \S4.1.
At first glance, this seems to slow the algorithm down, as~$\Omega(\gamma)$ operations are required for each component of the input.
Our experiments (refer to Figure~\ref{fig:time} below) confirm that the running time of~$\alg_*$ is particularly high compared to other alternatives.
Our research motivation is:
\begin{quote}
Can we accelerate this space-efficient Max Coverage algorithm, $\alg_*$, without significantly deteriorating space complexity or solution quality?
\end{quote}
One promising direction is to \emph{simplify} the subsampling process.
McGregor and Vu show that, with $\gamma = \lceil 2 c \varepsilon^{-2} k \log m \rceil$, an approximation factor of $1-1/e-\varepsilon$ is guaranteed with probability at least $1-{1}/{\text{poly}(m)}$. 
In the original version of~$\alg_*$,
this $\gamma$ parameter can easily exceed~$10^3$. So we would anticipate a thousand-fold reduction in throughput compared to a 
simpler, if theoretically less guaranteed, \emph{sampling scheme}, such as pairwise independent hashing.
Since we are designing a space-efficient algorithm, a pre-computed hash function table is infeasible.

\subsection{Our contributions}

Firstly, we show that the same space complexity and approximation quality can be achieved with~$\Theta(k \log m)$ independence (Corollary~\ref{cor:fstres}) instead of the original~$\Theta(\varepsilon^{-2} k \log m)$.

Secondly, we show how $F_0$-sketching can be avoided (Lemma~\ref{lemma:nosk}). This slightly reduces the probability of producing a $1-1/e-\varepsilon$ approximation, although not asymptotically. 

Finally, our experimental results demonstrate the efficiency and quality of our generalized algorithm, $\alg_\gamma'$. We observe that reducing the independence factor does not lead to significantly worse solutions. In particular, the algorithm still works with only pairwise independence. This leads to a significant speed-up, from $10\times$ to more than $1000\times$ for $k \geq 100$ (Figure \ref{fig:time}), while maintaining the same space complexity\footnote{Actually, removing $F_0$-sketching and reducing the independence factor strictly reduces the space complexity, although not asymptotically.} as the original algorithm of McGregor and Vu~\cite{McGregor2018}. Besides, we show that for reasonable values ($<0.27$) of~$\varepsilon$, our algorithm returns consistently better solutions than comparator streaming algorithms (Figure~\ref{fig:cov}).

\subsection{Related Work}

\textbf{Thresholding.}\label{sec:thresholding} Before surveying the algorithms for Max Coverage, we set out one of the important algorithmic frameworks.
Several algorithms invoke a thresholding technique, first applied to Set Cover by Cormode et al.~\cite{Cormode2010}.
It \emph{relaxes} the notion of greedy algorithm, and calculates a
near-greedy solution.
Instead of \emph{searching} for the set whose contribution is~$R^* = \max_j |S_j \setminus C|$,
a thresholding algorithm might add a set~$S_i$ to the solution if its contribution is at least~$\alpha R^*$, where $\alpha \in [0, 1]$ describes the \emph{greediness} of the thresholding algorithm. Applying this principle repeatedly results in a solution whose coverage is~$\alpha (1-1/e)$ fraction of the optimum coverage.

Now the guarantee of~$\alpha R^*$ contribution arises from a multi-pass approach to the stream. In pass~$j$, all sets with contribution at least~$r$ are added, then in pass~$j+1$, all sets with contribution at least~$\alpha r$ are added.
Since a set's contribution can only decrease as (other) sets are added to~$I$, with this approach, we only add a set if its contribution is~$\alpha R^*$.

\textbf{Prior art.} 
There are several existing streaming algorithms for the Max Coverage problem, which we summarize in Table~\ref{tab:streaming-alg}. Badanidiyuru et al.~\cite{Badanidiyuru2014} presented a generic algorithm for maximizing submodular functions on a stream, which can be adapted to Max Coverage. This is a one-pass thresholding algorithm, somewhat similar to $\alg_*$, that \textit{guesses} the optimal coverage size. Yu and Yuan~\cite{Yu2013} developed an algorithm that creates a specific ordering $(\Tilde{S}_1, \dots, \Tilde{S}_m)$ of the entire collection of sets $\{S_1, \dots, S_m\}$ such that for all $k$, $(\Tilde{S}_1, \dots, \Tilde{S}_k)$ is a solution of the \textit{Maximum-$k$-Coverage}.
Saha and Getoor~\cite{Saha2009}, who pioneered set streaming, took a \emph{swapping} approach.
A putative solution of~$k$ sets is stored, and sets in the putative solution can be replaced by new sets in the stream depending on the number of items uniquely covered by sets in the putative solution.
More recently, Bateni et al.~\cite{bateni2017almost} used a sketching technique and they almost match the optimal approximation factor of~$1-1/e$. This is an algorithm designed for the edge-arrival streaming model,
but can be adapted to the set streaming model with a space complexity independent to the size of the universe.
Norouzi-Fard et al.~\cite{NorouziFard2018}, in the continuation of Badanidiyuru et al.~\cite{Badanidiyuru2014}, presented a $2$-pass and a multi-pass approach to maximize a submodular function on a stream. 
Developed at a similar time, McGregor and Vu~\cite{McGregor2018} presented two polynomial-time algorithms that achieve the same approximation factor of $1-1/e-\varepsilon$: one taking a single pass, the other, $\alg_*$, taking multiple passes. 
The algorithms developed by McGregor and Vu~\cite{McGregor2018} are \textit{thresholding} algorithms.

\textbf{Sampling.} 
Sampling via hashing is a key component of many streaming algorithms. Relaxing the independence requirement for hash functions was explored in the context of $\ell_0$-samplers: Cormode and Firmani~\cite{l-sampl} invoked $\gamma$-independent hash functions. They showed theoretical bounds on~$\gamma$ to guarantee the probability of sampling a non-zero coordinate. In addition, their experimental results suggest that \textit{constant}-independence hashing schemes produce similar successful sampling rate to \textit{linear}-independent hash functions, while being significantly more efficient to compute.

Furthermore, some theoretical results \cite{inde-hash} show that many strong guarantees generally associated with \textit{high}-independence families of hash functions can be achieved with simpler hashing schemes.
Tabulation hashing~\cite{inde-hash}, for example, is not even $4$-independent, but manages to implement $\gamma$-independent hash function based algorithms, such as \textit{Cuckoo Hashing}.
P\v{a}tra{\c s}cu and Thorup \cite{inde-hash} also prove Chernoff-type inequalities with relaxed assumptions on the independence of the random variables.

\begin{table}
\caption{Streaming algorithms for Maximum Coverage. We focus on
the~$o(\min\{m,n\})$-space~algorithm, $\alg_*$.}
\label{tab:streaming-alg}
\begin{center}
\begin{tabular}{|l l | c c c |} 
     \hline
     \textbf{Author} & \textbf{Name} & \textbf{Passes} & \textbf{Space} & \textbf{Approx.} \\ 
     \hline
     \rule{0pt}{3ex} Badanidiyuru et al.~\cite{Badanidiyuru2014} &  $\mathsf{BMKK}$ & 1 & $\tilde{\mathcal{O}}(\varepsilon^{-1} n)$ & $1/2-\varepsilon$\\
     \hline
     \rule{0pt}{3ex} Yu and Yuan \cite{Yu2013} & & 1 & $\tilde{\mathcal{O}}(n)$ & $\sim 0.3$\\
     \hline
     \rule{0pt}{3ex} Saha and Getoor \cite{Saha2009} & $\mathsf{SG}$ & 1 & $\tilde{\mathcal{O}}(k n)$ & $1/4$\\
     \hline
     \rule{0pt}{3ex} Bateni et al. \cite{bateni2017almost} & & 1 & $\tilde{\mathcal{O}}(\varepsilon^{-3} m)$ & $1-1/e-\varepsilon$\\
     \hline
     \rule{0pt}{3ex} Norouzi-Fard et al. \cite{NorouziFard2018} & $\mathsf{2P}$ & $2$ & $\tilde{\mathcal{O}}(\varepsilon^{-1} n)$ & $5/9 - \varepsilon$\\
     \hline
     \rule{0pt}{3ex} Norouzi-Fard et al. \cite{NorouziFard2018} & & $\mathcal{O}(\varepsilon^{-1})$ & $\tilde{\mathcal{O}}(\varepsilon^{-1} n)$ & $1-1/e - \varepsilon$\\
     \hline
     \rule{0pt}{3ex} McGregor and Vu \cite{McGregor2018} & $\mathsf{OP}$ & 1 & $\tilde{\mathcal{O}}(\varepsilon^{-2} m)$ & $1-1/e-\varepsilon$\\
     \hline
     \rule{0pt}{3ex} McGregor and Vu \cite{McGregor2018} & $\alg_*$ & $\mathcal{O}(\varepsilon^{-1})$ & $\tilde{\mathcal{O}}(\varepsilon^{-2} k)$ & $1-1/e-\varepsilon$\\
     \hline
\end{tabular}

\end{center}
\end{table}

\section{Tools} 
\label{sec:background}

   The Introduction includes most of our notation; in addition, we let $I_\text{OPT}$ be
   an optimal solution and $\text{OPT}$ the size of the optimal coverage
   $\left| \bigcup_{i \in I_\text{OPT}} S_i \right|$.

\subsection{Subsampling}

\begin{definition}[subsampling]
Given $\family$, $\universe$, and hash function $h: \mathcal{U} \rightarrow \{0,1\}$,
the subsampled universe is~$\mathcal{U'} = \{ x \in \mathcal{U} \mid h(x) = 1 \}$,
with subsampled sets~$S' = S \cap \mathcal{U'}$ for every~$S \in \family$.
\end{definition}
Instead of computing with respect to universe $\mathcal{U}$,
algorithm $\alg_*$ focuses on~$\mathcal{U'} \subset \mathcal{U}$,
and tracks only the subsampled coverage $C'=\cup_{i \in I} S_i'$.
The size of the optimal coverage of~$\mathcal{U'}$, by a subfamily of~$k$ sets
from~\family, is henceforth called~$\text{OPT}'$.

\begin{remark*}
    The value~$\text{OPT}' = \max_{|J| = k} \left| \bigcup_{i \in J} S_i' \right|$ is not necessarily the same as the size of the union of the subsampled sets in the optimal coverage of~$\universe$, i.e., $\left| \bigcup_{i \in I_\text{OPT}} S_i' \right|$.
\end{remark*}

\begin{definition}
\label{prop:hash-functions}
Let $\gamma, v, p \in \mathbb{N}$ such that $p > |\mathcal{U}|$:
\[ \mathcal{H}_{\gamma, v} = \left\{ x \longmapsto \sum_{i=0}^{\gamma-1} a_i x^i \bmod p\  \bmod v \mid 0 \leq a_i < p \right\} \]
is a family of hash functions $\mathcal{H}_{\gamma, v} \subset \left\{ f: \mathcal{U} \rightarrow [v-1] \right\}$
\end{definition}

Such a family has the property of being \textit{$\gamma$-independent}\footnote{Different definitions exist; our definition of~$\gamma$-independent is stated in the Appendix.}. Evaluating a hash function $f \in \mathcal{H}_{\gamma, v}$ takes~$\Theta(\gamma)$ operations, including expensive \emph{modulo} operations, but these can be accelerated using the overflow mechanism on unsigned integer types. $\mathcal{H}_{\gamma, v}$ are the families of hash functions used in $\alg_*$.

\subsection{Sketching}
To estimate the size of a set, McGregor and Vu invoke~$F_0$ sketching.
\begin{theorem}[$F_0$-sketching \cite{Cormode2003}]
\label{th:sketching}
    Given a stream~$s$, there exists a data structure, $\mathcal{M}(s)$, that requires $\mathcal{O}(\varepsilon^{-2} \log \delta^{-1})$ space and, with probability $1-\delta$, returns the number of distinct elements in~$s$ within multiplicative factor $1\pm\varepsilon$. Processing each new element takes $\mathcal{O}(\varepsilon^{-2} \log \delta^{-1})$ time, the same time as finding the number of distinct elements.
\end{theorem}

\subsection{Thresholding on the sampled universe}
The core of $\alg_*$~is thresholding and subsampling.
The solution, $I$, and the associated subsampled
coverage $C'=\cup_{i \in I} S'_i$ are built incrementally, as new sets arrive in the stream and are selected. 
Given a threshold,~$r$, the selection rule for set $S_i$ is: 
\begin{equation}
\text{If } |S'_i \setminus C'| \geq r\text{, then } I \leftarrow I \cup \{i\} \text{ and }C' \leftarrow C' \cup S_i'\,,
\label{eqn:threshold}
\end{equation}
where $|S_i' \setminus C'|$ is called the \textit{contribution} of $S_i'$ -- from the context, it is clear this is in the sampled universe.
In choosing the sequence of thresholds there is a trade-off~\cite{chakrabarti2016incidence}: the larger the threshold, the higher the solution quality, but the more passes.

\section{Low-space Streaming Algorithm} 
\label{sec:algo}
In this section, we describe in detail $\alg_*$~developed by McGregor and Vu~\cite{McGregor2018}, which solves Max Cover in sublinear space with a respectable approximation factor.
Algorithm $\alg_*$~depends on two variables: 
\begin{itemize}
    \item $v$, an estimate of the optimal coverage, $\text{OPT}$; and
    \item $\lambda$, an estimate of the optimal coverage on the subsampled universe, $\text{OPT}'$.
\end{itemize}
These variables determine the probability of subsampling an element, and the initial value of the threshold, $r$, as applied above~\eqref{eqn:threshold}. The subsampling hash function is implemented as $h(x)=\mathds{1}_{f(x) < \lambda}$ where $f \in \mathcal{H}_{\lceil 2\lambda \rceil, v}$, hence the probability an item is subsampled is~$\lambda/v$. 
The threshold,~$r$, is initially set to $2(1+\varepsilon)\lambda/k$ and after each pass,~$r$ decreases by a factor~$1+\varepsilon$.
McGregor and Vu~\cite{McGregor2018} showed that if
\begin{equation}\lambda = c \varepsilon^{-2} k \log m\,, \qquad \text{with $c \geq 60$\,, and $\qquad \text{OPT}/2 \leq v \leq \text{OPT}$\,,}\label{eqn:lambda-def}
\end{equation}
then this thresholding procedure, which we call $\tp$, gives a $1-1/e-\varepsilon$ approximation using $\Tilde{\mathcal{O}}(\varepsilon^{-2} k)$ space with probability at least $1-1/\text{poly}(m)$.

\subsection{Guessing}

Algorithm $\alg_*$~relies on a reasonable estimate of $\text{OPT}$: a $v$ such that $\text{OPT}/2 \leq v \leq \text{OPT}$.
Of course, we do not know~$\text{OPT}$ in advance! The algorithm naively finds the right value for $v$ by executing $\tp_v$ for different values of $v$, called \textit{guesses}, in parallel.
Denote by $v_g$ the $g^\text{th}$ guess.
To reduce the number of guesses, we assume the maximum set size, which we call $||\mathcal{\family}||_\infty$, is known.
This assumption requires only one additional pass through the set stream, $\stream$: the \emph{asymptotic} number of passes is unchanged.
Hence the guesses for~$v$ can be restricted to all the values $v_g = 2^{g-2} ||\mathcal{S}||_\infty$, with~$g \geq 1$, smaller than $k ||\mathcal{S}||_\infty$.
These \emph{parallel} instantiations increase the running time and space by factor~$\log_2 k$: there are separate copies of variables~$I$, $C'$ and $h$ (the subsampling hash function) for each guess: these variables for guess~$v_g$ are $I_g$, $C_g'$ and $h_g$.
Now~$I$,~$C$ and~$C'$ refer to the variable associated with the output of the algorithm.

\begin{algorithm}[t]
\small
\caption{Algorithm $\alg_\gamma(\mathcal{S}, k, \varepsilon, ||S||_\infty)$}
\label{alg:withoutguess}
\DontPrintSemicolon
\nonl\Begin{
    \textbf{\textcolor{blue}{/* Initialise the guesses */}}\;
    $V \leftarrow \{ 2^{g-1} ||S||_\infty \leq \min(n, k ||S||_\infty), \ g \in \mathbb{N} \}$ \;
    Duplicates each variable $|V|$ times: $h$, $I$, $C'$, $\mathcal{M}$ and $\texttt{active}$\; 
    $r \leftarrow 2(1+\varepsilon)\lambda/k$\;
    \textbf{\textcolor{blue}{/* Multiple passes */}}\;
    \For{$p \gets 1$ \textbf{to} $1 + \lceil \log_{1+\varepsilon}(4e) \rceil$}{
        \textbf{\textcolor{blue}{/* One pass */}}\;
        \For{$S_i \in \stream$ \textbf{(stream)}}{
            \textbf{\textcolor{blue}{/* Iterate over the guesses */}}\;
            \For{$g \gets 0,\dots, |V|-1$}{
                $S_i' \leftarrow$ Subsample $S_i$ with $h_g$\;
                \label{line:subsample}
                $R_i \leftarrow S_i' \setminus C_g'$\ \textbf{\textcolor{blue}{/* Contribution */}}\;
                \label{line:contribution}
                \textbf{\textcolor{blue}{/* Check the \textit{bad guess} condition */}}\;
                \If{$|C_g'| + |R_i'| > 2(1+\varepsilon)\lambda$}{
                    \label{line:space-check}
                    $\texttt{active}_g \leftarrow$ \texttt{false}\;
                }
                \textbf{\textcolor{blue}{/* Thresholding procedure */}}\;
                \If{$\texttt{active}_g$ \textbf{and} $|I_g| < k$ \textbf{and} $|R_i| \geq r$}{
                    update $\mathcal{M}_g$ with $S_i$\;
                    $C_g' \leftarrow C_g' \cup R_i$\;
                    \label{line:update-cov}
                    $I_g \leftarrow I_g \cup \{i\}$\;
                    \label{line:update-sketch}
                }
            }
        }
        \textbf{\textcolor{blue}{/* Update the threshold */}}\;
        $r \leftarrow r / (1+\varepsilon)$\;
    }
    \textbf{\textcolor{blue}{/* Find the best coverage among the potentially correct guesses */}}\;
    $s \leftarrow \underset{\texttt{active}_g}{\mathrm{argmax}} \ \{|\mathcal{M}_g|\} $\;
    \label{line:select-guess}
    \Return $I_s$\; 
}
\end{algorithm}

\paragraph*{Which is the right guess?}
This guessing method begs the question: how do we detect the \emph{right} guess?
Also, $\alg_*$ is only guaranteed to \emph{work} under the condition $\text{OPT}/2 \leq v \leq \text{OPT}$. Some instances, with a \emph{wrong} guess, might necessitate more space than the bound $\tilde{\mathcal{O}}(\varepsilon^{-2} k)$. McGregor and Vu introduce two mechanisms to deal with these questions. 

First, the right guess is found by estimating the (non-subsampled) coverage of \universe associated with each guess:
the biggest coverage is considered the right guess.
However, only the subsampled coverages, of $\universe'$, are calculated. To resolve this, McGregor and Vu adopt \textit{$F_0$-sketching}, see Theorem~\ref{th:sketching}, which approximates the number of distinct elements in a collection of sets using less space than the collection itself.
More particularly, in addition to the subsampled coverage, $C_g'$, for each~$v_g$, $\alg_*$ maintains a \textit{sketch} $\mathcal{M}_g$ of the coverage in $\tilde{\mathcal{O}}(\varepsilon^{-2})$ space. Each time a set is selected, $\mathcal{M}_g$ is updated accordingly.
Once all the instances $\tp_g$ are completed, the sketches~$\{\mathcal{M}_g\}$  determine which guess produced the biggest coverage. We still need the subsampled coverages, $\{C_g'\}$, for calculating the set's contributions: $\{S_i \setminus C_g'\}$.  The $F_0$-sketches are too inefficient to be queried that often.

Second, the space complexity never exceeds $\tilde{\mathcal{O}}(\varepsilon^{-2} k)$ due to a consequence of Corollary~9 of McGregor and Vu~\cite{McGregor2018}: if $v$ is the right guess then $\text{OPT}' \leq 2 (1+\varepsilon) \lambda$. Thus, if the subsample coverage $C_g'$ of an instance $\tp_g$ exceeds $2 (1+\varepsilon) \lambda$, the associated guess is necessarily wrong and this instance can be terminated (Line~\ref{line:space-check} in Algorithm~\ref{alg:withoutguess}). Hence every instance runs in space~$\mathcal{O}(\lambda) = \tilde{\mathcal{O}}(\varepsilon^{-2} k)$. Thus, $\lambda$ can be referred as the \textit{space budget} of the algorithm.

\vspace{-10pt}
\subsection{Properties}
\label{sec:properties}

Algorithm~\ref{alg:withoutguess} is our generalisation of $\alg_*$, which we call~$\alg_\gamma$.
The independence factor, $\gamma$, is not fixed, but is instead a parameter that influences the implementation of the subsampling hash functions $\{h_g\}$.
The original algorithm, $\alg_*$, has an independence factor of $\lceil 2 \lambda \rceil$: \textbf{so $\alg_* = \alg_{\lceil 2 \lambda \rceil}$ in our generalization}.

McGregor and Vu~\cite{McGregor2018} showed that $\alg_*$ has space complexity $\tilde{\mathcal{O}}(\varepsilon^{-2} k)$
and with probability at least $1-\tfrac{1}{m^{10k}}$ produces a $1-1/e-\varepsilon$ approximation. 
As our focus is improving run time, with only a small trade-off in the other properties, we first dissect the run time.
The coverages $C_g'$ are implemented with hash tables in order to efficiently compute $R_i = S_i' \setminus C_g'$.
If the $g^\text{th}$ guess is reading the $i^\text{th}$ set then:
\begin{itemize}
    \item[Line \ref{line:subsample}:] Subsampling $S_i$:  $\mathcal{O}(\gamma)|S_i|$ time (evaluate degree-$\mathcal{O}(\gamma)$ polynomial for each element in $S_i$)
    \item[Line \ref{line:contribution}:] Computing $R_i$: $\mathcal{O}(|S_i'|) \subset \mathcal{O}(|S_i|)$ time
    \item[Line \ref{line:update-cov}:] Updating $C_g'$: $\mathcal{O}(|R_i|) \subset \mathcal{O}(|S_i|)$ time
    \item[Line \ref{line:update-sketch}:] Updating $\mathcal{M}_g$:  $\mathcal{O}(\varepsilon^{-2} \log n) |S_i|$ time (Theorem~\ref{th:sketching})
\end{itemize}
\noindent
Therefore, the expected time complexity, $T_\gamma$, of $\alg_\gamma$ is
\begin{align}
    T_\gamma & = \underbrace{\mathcal{O}(\log k)}_{\text{guesses}}
    \left(
        \underbrace{\mathcal{O}(\varepsilon^{-1})}_{\text{passes}} 
        \cdot \underbrace{\mathcal{O}\left( \sum_{i=1}^m \gamma |S_i| \right)}_{\text{subsampling}} + 
        \underbrace{\mathcal{O}\left( \sum_{i \in I} \varepsilon^{-2} |S_i| \log n \right)}_{F_0\text{-sketching}} 
    \right) \notag\\
    & = \mathcal{O}\left( \varepsilon^{-1} \gamma m |\overline{\mathcal{S}}| \log k + \varepsilon^{-2} k |\overline{\mathcal{C}}| \log n \log k \right) \label{eqn:Tgamma}
 \end{align}

\begin{note*}
$|\overline{\mathcal{S}}| = \tfrac{1}{m}\sum_i |S_i|$ is the average set size over the entire stream,~$\mathcal{S}$, while
    $|\overline{\mathcal{C}}| =\tfrac{1}{k} \sum_{i \in I} |S_i|$ is the average set size over the selected sets (in~$I$).
\end{note*}
Therefore, $\alg_* = \alg_{\lceil 2 \lambda \rceil}$ has expected time complexity of \begin{equation}T_{\lceil 2 \lambda \rceil} = \mathcal{O}\left( \varepsilon^{-3} k m |\overline{\mathcal{S}}| \log m \log k + \varepsilon^{-2} k |\overline{\mathcal{C}}| \log n \log k \right)\,.\label{eqn:Ttwolambda}\end{equation}

Regarding the space complexity for $\gamma < \lceil 2 \lambda \rceil$, it remains (asymptotically) the same. Indeed the cost for storing a $\gamma$-independent hash function is $\mathcal{O}(\gamma) \subset \Tilde{\mathcal{O}}(\varepsilon^{-2} k)$.

Interestingly,~$\alg_\gamma$ does not guarantee the solution returned actually has~$k$ sets.
Given that~$|I| \leq k$, we can simply append $k-|I|$ random indices to the returned solution. However, the goal of this paper is to assess the probabilistic nature of the algorithm that arises from the $\gamma$-independent hash functions. Therefore, in our experiments in \S\ref{sec:experimentation}, we do not alter the returned solution given by $\alg_\gamma$.

\section{Accelerating the Algorithm}
\label{sec:lowering-independence-factor}

Algorithm $\alg_\gamma$ is a breakthrough: it runs in sublinear space.
As we surmise from the expression~\eqref{eqn:Tgamma} for~$T_\gamma$, however, the high independence factor,~$\gamma$, induces a significant bottleneck.
Our experiments in \S\ref{sec:experimentation} validate this conjecture.
In this section we demonstrate how to improve the running time without too much sacrifice in the other properties.
Indeed, when lowering the independence factor,~$\gamma$, we maintain the space complexity and the number of passes.

McGregor and Vu \cite{McGregor2018} showed that if $\text{OPT}/2 \leq v$ and $\big| \ |C'| - p|C| \ \big| < \varepsilon v p$ then the $I$ is a $1-1/e-\varepsilon$ approximation.
Assuming condition $\text{OPT}/2 \leq v \leq \text{OPT}$ is met, i.e., we have the right guess for~$v$, $\mathbb{P}(E_I)$ is a lower bound 
on the probability of producing a $1-1/e-\varepsilon$ approximation.
where~$E_J$ is the event $\mathbb{P}\left(\,\big| \ |\cup_{j\in J} S_j'| - p|\cup_{j\in J} S_j| \ \big| < \varepsilon v p \,\right)$. Therefore, the goal is to estimate $\mathbb{P}(E_I)$ for independence factor~$\gamma$ smaller than $\lceil 2\lambda \rceil$. Note that $E_J$ is the event associated to an arbitrary set of indices $J$ while $E_I$ is associated with the output of $\alg_\gamma$, $J$ is a constant while $I$ is a random variable. The goal is to estimate $\mathbb{P}(E_I)$ but it is much easier to estimate $\mathbb{P}(E_J)$ for an arbitrary set of indices $J$. Therefore, the following results focus on $\mathbb{P}(E_J)$ but notice that $\bigcap_{|J| = k} E_J \subset E_I$ so:

\[ \mathbb{P}\left(\overline{E_I}\right) \leq \mathbb{P}\left( \bigcup_{|J| = k} \overline{E_J} \right) \leq \sum_{|J| = k} \mathbb{P}(\overline{E_J}) \]

We provide proofs of several of the following results in the Appendix.

\subsection{Optimising the Independence Factory}

\textbf{Recalculating~\texorpdfstring{$c$}{c}.}
The first optimisation is actually an observation about the constant~$c$ in the definition of $\lambda = c \varepsilon^{-2} k \log m$.
Indeed, the independence factor $\gamma = \lceil 2 \lambda \rceil$ depends on $c$:
the authors state ``Let~$c$ be some sufficiently large constant.''
A close inspection of the proof of their Lemma 8 reveals that $c$ is set to a value bigger than $60$ but we can find a smaller value that still gives a $1-1/\text{poly}(m)$ probability of success. In order to express an explicit upper bound for the failure probability we can use an intermediate result from the proof of Lemma 8: $\mathbb{P}(\overline{E_J}) \leq e/m^{ck/6}$.
Noticing that $\bigcap_{|J| = k} E_J \subset E_I$ we find

\[ \mathbb{P}\left(\overline{E_I}\right) \leq \sum_{|J| = k} \mathbb{P}(\overline{E_J}) \leq \binom{m}{k} \frac{e}{m^{ck/6}} \leq \frac{e}{k! \ m^{(c/6-1)k}}\,, \]

whence we conclude that ~$c=6$ is the smallest reasonable value to be sure the upper bound is~$o(1)$.
This reduction in~$c$ does not reduce the asymptotic time complexity of $\alg_\gamma$, but in practice it reduces the independence factor by a factor~$10$ so this is still a $10 \times$ speed up compared to the original $c \geq 60$.

\textbf{Reducing~\texorpdfstring{$\gamma$}{Y}.}
In order to reduce further the independence factor $\gamma$, we express $\mathbb{P}(E_J)$ with respect to~$\gamma$. To that end, we use the following concentration bound, from the same paper cited by McGregor and Vu.

\begin{theorem}[Schmidt et al.~\cite{Schmidt1995}]
\label{th:gen}
Let $X_1, \dots, X_n$ be $\gamma$-wise independent r.v.s, $X = \sum_{i=1}^n X_i$ and $\mu = \mathbb{E}(X)$. If $X_i \in \{0, 1\}$ and $\gamma \leq \lfloor \min(\delta, \delta^2) \mu e^{-1/3} \rfloor$ then $\mathbb{P}(| X - \mu | \geq \delta \mu) \leq e^{-\lfloor \gamma / 2 \rfloor}$.
\end{theorem}
Notice that $|C'|=\sum_{i=1}^n X_i$ where $X_i = \mathds{1}_{i \in C} \mathds{1}_{h(i)=1} \in [0, 1]$; since $p=\lambda/v$ is the probability of subsampling an element, $\mathbb{P}(h(i) = 1)$, we have the following corollary:
\begin{corollary}
\label{cor:fstres}
If $\gamma \leq \lfloor \tfrac{c}{3} k \log m \rfloor$, then:
\[ \mathbb{P}\left(\overline{E_J} \right) \leq e^{-\lfloor \gamma / 2 \rfloor}\,. \]
\end{corollary} 

Consequently, by setting $\gamma = \lfloor \tfrac{c}{3} k \log m \rfloor = \mathcal{O}(\varepsilon^2 \lambda)$, compared to the original $\mathcal{O}(\lambda)$, we keep the same approximation guarantees as McGregor and Vu~\cite{McGregor2018}:
\[ \mathbb{P}(\overline{E_J}) \leq e^{-\left\lfloor \left\lfloor \tfrac{c}{3} k \log m \right\rfloor / 2 \right\rfloor} = e^{-\left\lfloor \tfrac{c}{6} k \log m \right\rfloor} \leq e^{-\tfrac{c}{6} k \log m + 1} = \frac{e}{m^{ck/6}}\,. \]

\subsection{Removing \texorpdfstring{$\bm{F_0}$}{F0}-sketching.}
First, it should be noted that McGregor and Vu showed that $\alg_*$ produces a $1-1/e-\delta(\varepsilon)$ approximation of the optimal coverage where $\delta(\varepsilon)=\varepsilon(3-1/e-\varepsilon)\leq 2.6\varepsilon$. Asymptotically, the statement of McGregor and Vu is right because the algorithm can simply start by dividing $\varepsilon$ by $3$ and it would indeed produce a $1-1/e-\varepsilon$ approximation. Nevertheless, such a modification would result in a significant slowdown ($\times 3$ to $\times 27$ depending on the independence factor). In the \S\ref{sec:experimentation} experiments, we assess the approximation quality relatively to the actual theoretical bound $1-1/e-\delta(\varepsilon)$. Finally, thanks to the following result, we adapt $\alg_\gamma$ so that $F_0$-sketching is not needed. 

Let $\alg_\gamma'$ be the algorithm that replaces line~\ref{line:select-guess} in Algorithm~\ref{alg:withoutguess} with Algorithm~\ref{alg:findguess}. The selected guess is the biggest active guess,~$s$, such that $|C_s'| \geq (1-\varepsilon)(1-1/e-\varepsilon)\lambda$. We thus conclude $\alg_\gamma'$ is correct for $\gamma \geq \lfloor \tfrac{c}{3} k \log m \rfloor$.
\begin{algorithm}[t]
\small
\caption{Procedure \texttt{FindGuess}}
\label{alg:findguess}
\DontPrintSemicolon
\nonl$s \leftarrow |V|-1$\;
\nonl\While{$|C_s'| < (1-\varepsilon)(1-1/e - \varepsilon)\lambda$ \textbf{or} $\neg\texttt{active}_s$}{
    \nonl$s \leftarrow s - 1$\;
}
\end{algorithm}

\begin{lemma}
\label{lemma:nosk}
Let $v$ be some guess in $\alg_\gamma'$ and let $C'$ be the final subsampled coverage associated with guess $v$. If 
$$  \big| \ |C'| - p|C| \ \big| < \varepsilon v p\text{; and }
    v > \text{OPT}\text{; and }
    (1-\varepsilon)(1-1/e-\varepsilon)\lambda \leq |C'|\,,
$$
then $|C| > (1-1/e-\delta(\varepsilon))\text{OPT}$.
\end{lemma}

\begin{proposition}
\label{prop:nosk}
    For $\gamma \geq \lfloor \tfrac{c}{3} k \log m \rfloor$, $\alg_\gamma'$ finds a $1-1/e-\delta(\varepsilon)$ approximation of the \textit{Maximum-$k$-Coverage} problem with probability at least $1 - 2 e/(k! m^{(c/6-1)k})$.
\end{proposition}

Since the~$F_0$-sketch is omitted, the time complexity is $T_\gamma' = \mathcal{O}( \varepsilon^{-1} \gamma m |\overline{\mathcal{S}}| \log k )$, while the space complexity is unchanged. With $\gamma = \lfloor \tfrac{c}{3} k \log m \rfloor$, $\alg_\gamma'$ has a time complexity of $T_{\lfloor \varepsilon^2 \lambda / 3 \rfloor}' = \mathcal{O}( \varepsilon^{-1} k m |\overline{\mathcal{S}}| \log k \log m)$,
which is at least~$\varepsilon^{-2}$ faster than expression~\eqref{eqn:Ttwolambda}.

\section{Experimentation}
\label{sec:experimentation}

In this section, we assess the performance of the  $\alg_\gamma'$ algorithm family on real-world datasets. We focus on four datasets,  summarized in Table~\ref{tab:datasets}:
\begin{itemize}
    \item \textit{SocialNet}\footnote{\url{https://snap.stanford.edu/data/com-Friendster.html}}
    represents a collection of individuals linked by a friendship relation.
    \item \textit{UKUnion} \cite{UKUnion} combines snapshots of webpages in the .uk domain taken over a 12-month period between June 2006 and May 2007.
    \item \textit{Webbase} \cite{Webbase} and \textit{Webdocs} \cite{goethals2003fimi} each represent a collection of interlinked websites.
\end{itemize}

\begin{table}[ht]
\caption{Real-world datasets. Hapax Legomena (HL) refers to the number of sets that contains an element which appears only in this set. The minimum set size and element frequency is~$1$.}
\label{tab:datasets}
\begin{center}
\begin{tabular}{|c | r r | r r r | r c r | c |} 
     \hline
     \multirow{2}{*}{\textbf{Dataset}} & \multicolumn{1}{c}{$\bm{n}$} & \multicolumn{1}{c|}{$\bm{m}$} & \multicolumn{3}{c|}{\textbf{Set size}} & \multicolumn{3}{c|}{\textbf{Element frequency}} & \multirow{2}{*}{\textbf{HL}} \\ 
      & $\times 10^6$ & $\times 10^6$ & \multicolumn{1}{c}{Max} & \multicolumn{1}{c}{Med} & \multicolumn{1}{c|}{Avg} & \multicolumn{1}{c}{Max} & \multicolumn{1}{c}{Med} & \multicolumn{1}{c|}{Avg} & \\
      \hline
      SocialNet & 65.0  & 37.6 & 3,615  & 12    & 48.10 & 4,223     & 6 & 27.80 & 24.0\% \\
      UKUnion   & 126.5   & 74.1 & 22,429 & 25    & 45.56 & 4,714,511 & 2 & 26.71 & 16.7\% \\
      Webbase   & 112.2   & 57.0 & 3,841  & 6     & 11.81 & 618,957   & 2 & 6.00  & 23.5\%  \\
      Webdocs   & 5.3  & 1.7 & 71,472 & 98    & 177.20 & 1,429,525 & 1 & 56.93 & 21.2\% \\
     \hline
\end{tabular}
\end{center}
\end{table}

$\alg_\gamma'$ is implemented\footnote{\url{https://github.com/caesiumCode/streaming-maximum-cover}} in C++20 and executed on \textit{Spartan} the high
performance computing system of The University of Melbourne. The CPU
model is the Intel(R) Xeon(R) Gold 6254 CPU @ 3.10GHz with a maximum
frequency of 4GHz. $\alg_\gamma'$ naturally implies a parallel algorithm that consists of performing the computation related to each guess in parallel. Nonetheless, we do not implement an actual parallel algorithm as it would require substantial effort in order to fine tune. Also, compared to the original algorithm, this approach does not change the number of guesses. Therefore, the potential speed-up of a parallel implementation would be the same for our algorithm $\alg_\gamma'$ and the original algorithm $\alg_*$.

\textbf{Assessing coverage.}
With original independence factor $\gamma = \lceil 2 \lambda \rceil$, $\alg_\gamma'$ can still take tens of hours on the biggest datasets. We thus introduce a new variant of the algorithm, the \textit{full sampling} variant, $\alg_{\text{fs}}'$. \textit{Full sampling} means there is no subsampling so $\alg_{\text{fs}}'$ is a deterministic algorithm where $\mathbb{P}(E_I) = 1$. It means that $\alg_{\text{fs}}'$ is fast and produces particularly good solutions (Figure \ref{fig:cov}). However, it has a space complexity of $\Tilde{\mathcal{O}}(n)$ so $\alg_{\text{fs}}'$ is just seen as a tool to assess the approximation quality of $\alg_2'$. 

\textbf{Setting~$c$.}
We only have theoretical guarantees for $c > 6$. However, we used large overestimates to derive the inequality. For that reason we set $c=1$ for the experiments.

\textbf{Suite of experiments.}
Algorithms $\algfs'$, $\algfull'$, $\algopt'$ and $\algpair'$ are executed on the four datasets, for $\varepsilon \in \{\tfrac{1}{2}, \tfrac{1}{4}, \tfrac{1}{8}\}$ and $k \in \{ 1, 2, 4, 8, 16, 32, 64, 128, 256 \}$. Although $\varepsilon=1/2$ is out of the theoretical consideration, because $1-1/e-\delta(1/2) < 0$, it presents an opportunity to observe how the algorithm behaves outside its theoretical scope: $\varepsilon < 0.267$.
For each Figure, in the main text, we only show the datasets representative of the variety of behaviors. The remaining components of each figure are in the Appendix.

\textbf{Comparator algorithms (refer to Table \ref{tab:streaming-alg}).}
The goal is to assess the trade-off between time, space and approximation quality, therefore we aim to compare $\alg_\gamma'$ with algorithms that perform relatively well in all three categories. For that reason, we implemented the comparator algorithms $\mathsf{SG}$, $\mathsf{BMKK}$, and $\mathsf{2P}$. The algorithm of Yu and Yuan~\cite{Yu2013} takes too much time and space, as it solves for all possible values of~$k$ at once. Also, as illustrated and explained in the Appendix (Figure~\ref{annex:cov-space-op}), the $\Tilde{\mathcal{O}}(\varepsilon^{-d} m)$-space algorithms consume too much space in comparison to the $\Tilde{\mathcal{O}}(\varepsilon^{-d} n)$ space algorithms. We run Algorithm~$\mathsf{2P}$ instead of its $(1-1/e-\varepsilon)$-approximation cousin~\cite{NorouziFard2018}: the latter is less effective than~$\mathsf{2P}$, empirically, while consuming the same space and taking more passes.

\subsection{Runtime Evaluation} 
\begin{figure}
    \centering

    \centerline{\includegraphics[width=\textwidth]{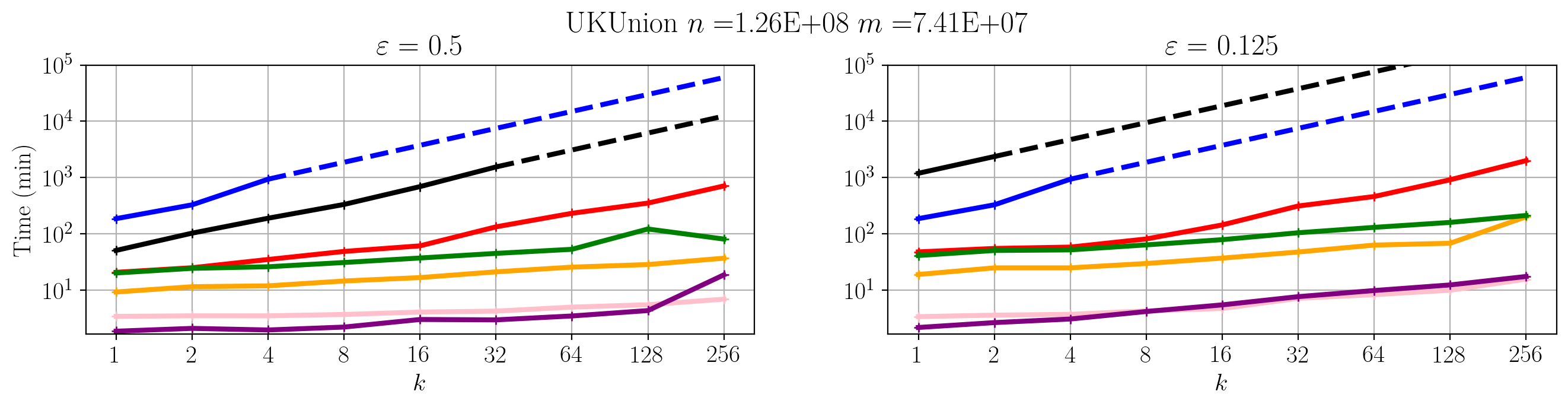}}
    \centerline{\includegraphics[width=\textwidth]{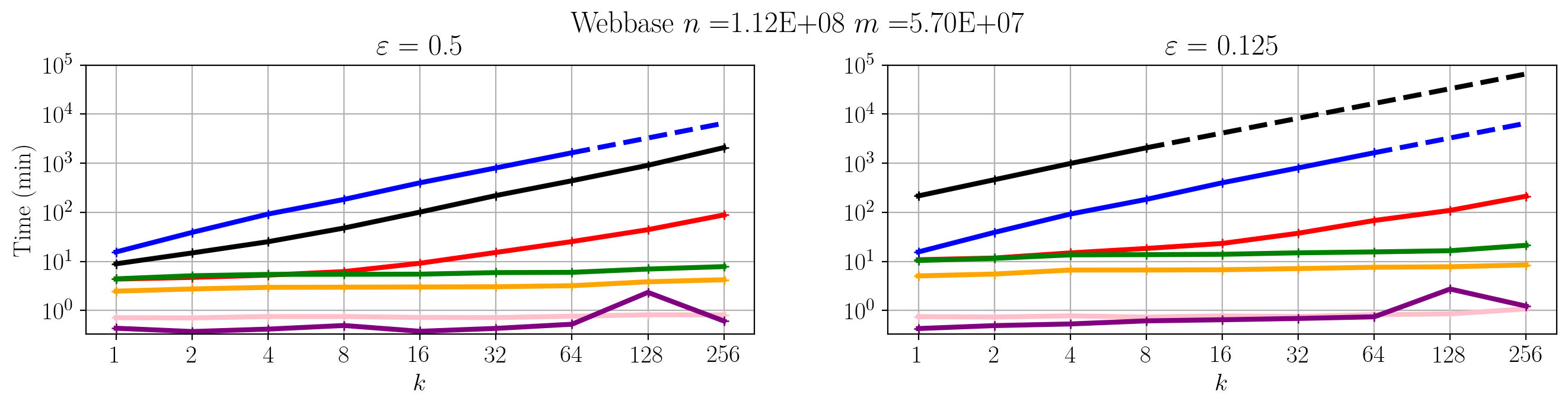}}

    \centerline{\includegraphics[width=\textwidth]{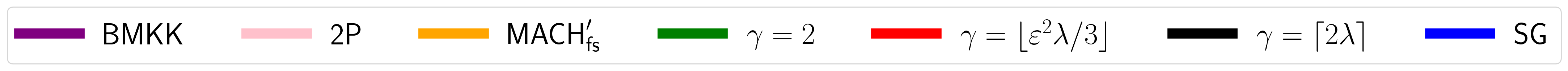}}
    \caption{Running times of the algorithms, demonstrated on \textit{Webbase}. Observe that for~$\varepsilon=1/8$, sticking with~$\gamma=\lceil 2\lambda \rceil$ leads to a particularly slow algorithm. On the other hand, up to~$k=8$, both $\gamma=\lceil \varepsilon^2\lambda/3 \rceil$ and~$\gamma=2$ are only $2$--$3$ times slower than full sampling. For larger values of~$k$, $\gamma=\lceil \varepsilon^2\lambda/3 \rceil$ becomes $8$--$10$ times slower. The missing values (dashed line) for $\gamma = \lceil 2 \lambda \rceil$ and $\mathsf{SG}$ are extrapolated, they refer to a running time that exceeds the time limit of~48 hours. 
    }
    \label{fig:time}
\end{figure}

Figure~\ref{fig:time} demonstrates the time saved by reducing the independence factor. $\algopt'$ is consistently faster than $\algfull'$ by an order of magnitude, while,  as~$k$ increases, $\algpair'$ widens its gap over $\algopt'$. In contrast to $\algfs'$, the time spent calculating hash function outputs for subsampling is clear. About half the running time of $\algpair'$ is about subsampling, while this proportion easily exceeds~99\% of the running time for $\algfull'$. Considering comparators, $\mathsf{BMKK}$ and $\mathsf{2P}$ are equally the fastest algorithms by a wide margin. Despite only one pass, $\mathsf{SG}$ is one of the slowest algorithms, along with $\algfull'$. The precise running times can be consulted in Table \ref{tab:time}.

\setlength{\tabcolsep}{0.05em}
\begin{table}
\caption{Summary of the running times in minutes of the algorithms (Figure \ref{fig:time}). An empty cell means the time exceeds 48 hours (2880 minutes).}
\label{tab:time}
\begin{center}
\begin{tabular}{| l | *{4}{S[table-column-width=4.4em]} | *{4}{S[table-column-width=4.4em]} |} 
    \hline
    \multicolumn{1}{|c|}{Dataset} & \multicolumn{4}{c|}{\textit{Webbase}} & \multicolumn{4}{c|}{\textit{UKUnion}}\\
    \multicolumn{1}{|c|}{$k$} & 4 & 16 & 64 & 256 & 4 & 16 & 64 & 256\\
    \hline
    \multicolumn{1}{|c}{} & \multicolumn{8}{c|}{$\varepsilon = 0.5$}\\
    \hline
$\algfs'$ & 4.5 & 4.4 & 4.7 & 9.2  & 17.1 & 22.3 & 38.1 & 59.7\\
$\algpair'$ & 8.5 & 8.8 & 9.3 & 10.8  & 35.7 & 48.3 & 77.2 & 134.5\\
$\algopt'$ & 8.0 & 14.5 & 38.3 & 135.9  & 40.8 & 90.3 & 338.3 & 1145.3\\
$\algfull'$ & 143.4 & 588.6 & 2609.1 & \textbf{-}  & 849.0 & \textbf{-} & \textbf{-} & \textbf{-}\\
$\mathsf{SG}$ & 92.1 & 399.3 & 1632.4 & \textbf{-}  & 936.7 & \textbf{-} & \textbf{-} & \textbf{-}\\
$\mathsf{BMKK}$ & 0.5 & 0.5 & 0.6 & 0.8  & 2.9 & 3.8 & 5.5 & 11.6\\
$\mathsf{2P}$ & 0.8 & 0.8 & 0.8 & 0.8  & 3.6 & 4.4 & 6.6 & 8.8\\
    \hline
    \multicolumn{1}{|c}{} & \multicolumn{8}{c|}{$\varepsilon = 0.125$}\\
    \hline
$\algfs'$ & 6.7 & 6.8 & 7.6 & 8.5  & 24.9 & 37.1 & 63.3 & 201.3\\
$\algpair'$ & 13.6 & 13.9 & 15.6 & 21.3  & 52.0 & 79.0 & 130.6 & 212.4\\
$\algopt'$ & 14.9 & 23.2 & 68.0 & 212.7  & 58.3 & 144.3 & 461.3 & 2002.5\\
$\algfull'$ & 987.4 & \textbf{-} & \textbf{-} & \textbf{-}  & \textbf{-} & \textbf{-} & \textbf{-} & \textbf{-}\\
$\mathsf{SG}$ & 92.1 & 399.3 & 1632.4 & \textbf{-}  & 936.7 & \textbf{-} & \textbf{-} & \textbf{-}\\
$\mathsf{BMKK}$ & 0.5 & 0.6 & 0.7 & 1.2  & 3.1 & 5.5 & 9.9 & 17.5\\
$\mathsf{2P}$ & 0.8 & 0.8 & 0.8 & 1.1  & 3.7 & 4.7 & 8.3 & 15.7\\
    \hline
\end{tabular} 
\end{center}
\end{table}

\subsection{Space Efficiency}
\begin{figure}
    \centering
    \centerline{\includegraphics[width=\textwidth]{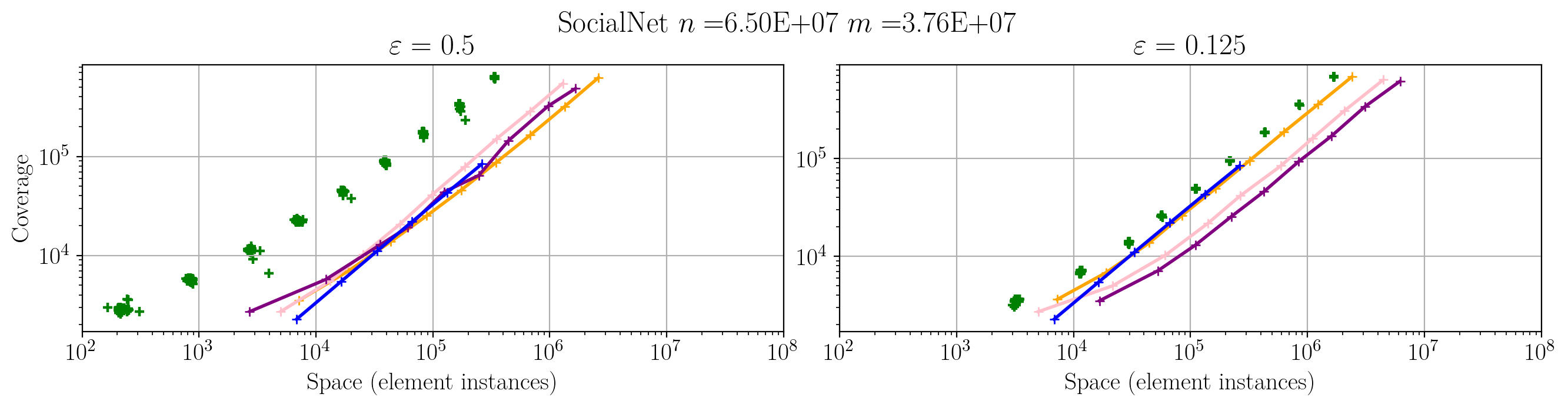}}
    \centerline{\includegraphics[width=\textwidth]{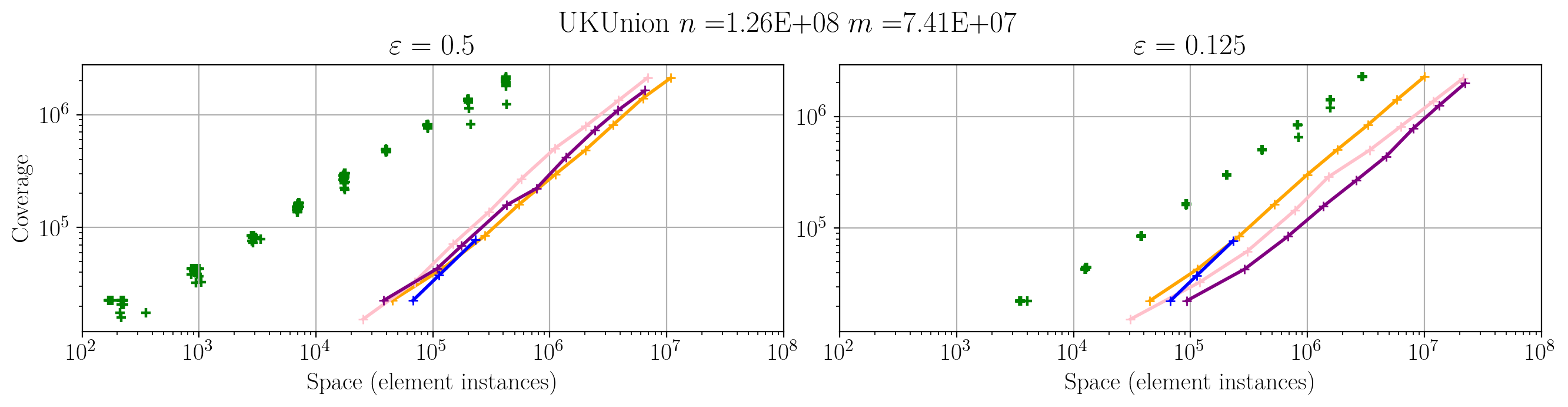}}

    \centerline{\includegraphics[width=\textwidth]{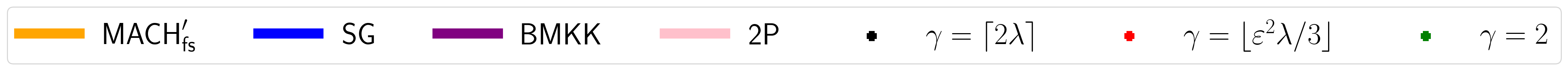}}
    \caption{Coverage versus space, for~$k \in \{1,2,4,8,16,32,64,128,256\}$. Every panel shows the advantage of the~$\alg'$ family for max coverage in streams. For~$\varepsilon=1/2$, the space advantage is at least ten-fold. Interestingly, as~$\varepsilon$ decreases, the space advantage drops for some datasets, but the coverage does not improve significantly, suggesting that a \textit{lightweight}~$\alg'$ approach, i.e., smaller~$\varepsilon^{-1}$, might be the most effective time-space-performance trade-off.}
    \label{fig:space-cov}
\end{figure}

To measure the space complexity of the different algorithms, we simply count the number of element instances stored by each algorithm. Figure~\ref{fig:space-cov} demonstrates how space efficient $\alg_\gamma'$ is compared with alternatives, as predicted by the sublinear asymptotic bound: $\Tilde{\mathcal{O}}(\varepsilon^{-2} k)$, seemingly independent of the coverage size, in practice as well as theory.

As stated earlier, the space complexity of $\algfs'$, $\mathsf{SG}$ and $\mathsf{BMKK}$ scales linearly with the coverage size of the solution. So when $\alg_\gamma'$ does not look so advantageous for \textit{SocialNet} when $\varepsilon = 0.125$, it is simply because the coverage is almost as small as the space budget of $\alg_\gamma'$. The coverage of \textit{UKUnion} is about 10 times bigger than \textit{SocialNet}, but the space consumption is about the same as \textit{SocialNet}.

\subsection{Estimating Approximation Quality} 
\begin{figure}
    \centering

    \centerline{\includegraphics[width=\textwidth]{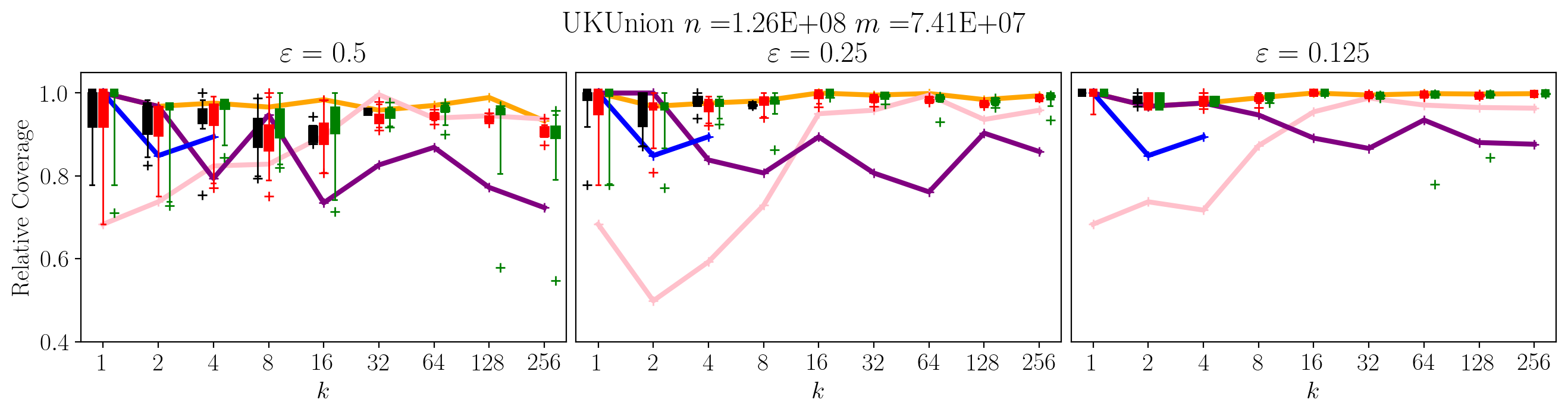}}
    
    \centerline{\includegraphics[width=\textwidth]{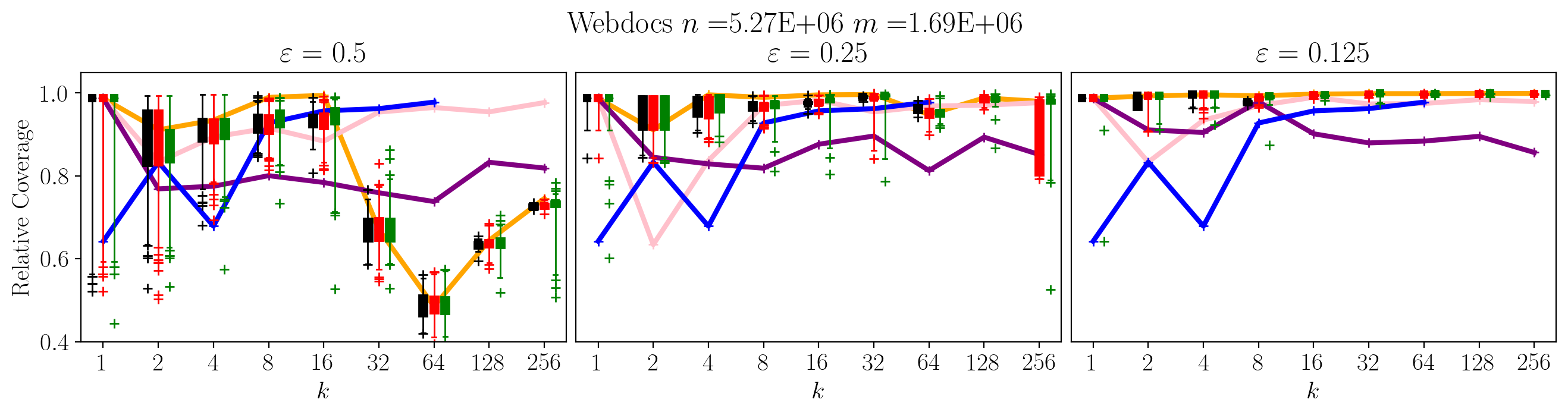}}
    \centerline{\includegraphics[width=\textwidth]{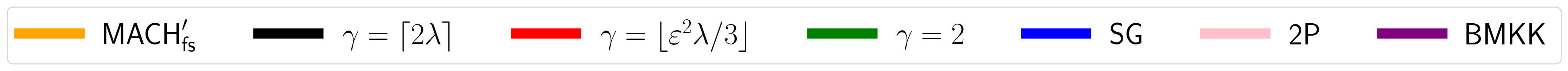}}
    \caption{Coverage of the algorithms relative to greedy coverage. Since they are randomized, there are box plots of coverage produced by~$\alg_\gamma'$.  
    The box plots show the 1\%, 25\%, 75\% and 99\% quantiles, hence the points below and above the box plots are in the first and last 1\%. For $\gamma = 2$ and $\gamma = \lfloor \varepsilon^2 \lambda /3 \rfloor$, each boxplot gathers~200 data points on average whereas for $\gamma = \lceil 2 \lambda \rceil$, each boxplot gathers~90 data points on average. Observe that the~$\alg'$ methods return excellent coverage except for values of~$k$ around~$64$ on the dataset \textit{Webdocs} when $\varepsilon = 1/2$. 
    }
    \label{fig:cov}
\end{figure}

The maximum set coverage problem is NP-Hard. Comparing the coverage size produced by $\alg_\gamma'$ with the optimal solution is infeasible at the scale of our datasets.
Since the greedy algorithm guarantees a $1-1/e$ approximation, and can be implemented, its coverage is our \emph{reference}. Moreover, even if the optimal solution, $\text{OPT}$, is unknown, the $1-1/e-\delta(\varepsilon)$ approximation of $\alg_\gamma'$ can be verified using the greedy algorithm thanks to the following implication:
${|C|}/{|G|} \geq \beta \ \Longrightarrow \ {|C|} /{\text{OPT}} \geq \beta (1-1/e)$,
where~$G$ returned by greedy and $C$ an arbitrary coverage. In particular, if ${|C|}/{|G|} \geq 1 - \delta(\varepsilon)/(1-1/e)$ then $|C|/{\text{OPT}} \geq 1-1/e-\delta(\varepsilon)$. 

Figure~\ref{fig:cov} demonstrates that for theoretically \textit{admissible} values of~$\varepsilon$, $\alg_\gamma'$ produces a coverage close to greedy coverage, and always within the $1-{\delta(\varepsilon)}/(1-1/e)$ limit. Regardless of~$\gamma$ value, it remains very close to the coverage produced by $\alg_{\text{fs}}'$. Even for pairwise independence, which is expected to produce worse solutions, there is no clear performance effectiveness difference compared higher independence. Additionally, even though $\alg_\gamma'$ is a randomized algorithm, no coverage has been observed to beat greedy.

sWe observe some rare events (< $1\%$) where the coverage is particularly small compared to $\alg_\text{fs}'$. Investigating these cases reveals that such solutions typically contain fewer than $k$ sets. If $\alg_\gamma'$ does not select the right guess, it tends to select a guess slightly bigger than the right one, which increases the threshold, therefore it does not have enough opportunity to select $k$ sets in $\mathcal{O}(\varepsilon^{-1})$ passes.

\section{Conclusions}

In this paper, we accelerate the sublinear-space approach to solving Maximum Coverage.
The algorithm $\alg_*$ of McGregor and Vu is hampered by a high-independence hash function.
We generalize their approach to produce~$\alg_\gamma$, so that
$\alg_* = \alg_{\lceil 2 \lambda \rceil}$
and then avoid $F_0$-sketches to obtain
$\alg_\gamma'$.
The space consumption is in~$\Tilde{\mathcal{O}}(\varepsilon^{-2} k)$ and the approximation factor is~$1-1/e-\delta(\varepsilon)$.

For reasonable values of~$\varepsilon$ ($\leq 0.25$), our algorithm, $\alg_\gamma'$, maintains the space efficiency and approximation quality of $\alg_* = \alg_{\lceil 2 \lambda \rceil}$. In experiments, it is several orders of magnitude faster. In practice, we find $\alg_2'$ presents the best trade-off between space complexity, time complexity and approximation quality.
Since~$\alg'_2$ is so efficient, we can run it several times with fresh randomness. This approach is more effective than executing~$\alg_\gamma'$ with a high independence factor. Although we avoided $F_0$-sketching in $\alg'_\gamma$, they could help compare independent instances of the fast $\alg_2'$.

We obtained several key results by carefully analyzing upper bounds on algorithm failure probability.
We expect this idea accelerates other lower-space streaming algorithms.

\clearpage

\clearpage

\appendix

\section[Definition of y-independent hash functions family]{Definition of $\gamma$-independent hash functions family}

\begin{definition}
The family of hash functions $\mathcal{H} \subset \left\{ f:\mathcal{U} \rightarrow Y \right\}$ is $\gamma$-independent iff for every~$\gamma$ distinct keys $x_1,\dots,x_\gamma \in \mathcal{U}$ and~$\gamma$ values $y_1,\dots,y_\gamma \in Y$, if we draw~$f$ uniformly  at random from~$\mathcal{H}$,
then the~$f(x_i)$ are independent uniform random variables, and
$\mathbb{P}\left[\cap_{i=1}^\gamma ( f(x_i) = y_i ) \ \right] = {1}/{|Y|^\gamma}$. 
\end{definition}

\section{Corollary \ref{cor:fstres} (proof)}

\begin{corollary5}
If $\gamma \leq \lfloor \tfrac{c}{3} k \log m \rfloor$, with~$I$,~$C$, and~$C'$ defined accordingly, then:
\[ \mathbb{P}\left(\ \big| \ |C'| - p|C| \ \big| \geq \varepsilon v p \ \right) \leq e^{-\lfloor \gamma / 2 \rfloor}\,. \]
\end{corollary5}

\begin{proof}
Consider~$I$ to be fixed, so the $\mathds{1}_{i\in C}$ factor is just a constant, with no randomness.
\[ \mathbb{E}(|C'|) = \sum_{i=1}^n \mathds{1}_{i\in C} \mathbb{E}(\mathds{1}_{h(i)=1}) = \sum_{i\in C} \mathbb{P}(h(i)=1) = p |C| \]
Let $\delta = \varepsilon v / |C|$ and $\mu = \mathbb{E}(|C'|)$, then 
$ \mathbb{P}\left( \ \big| \ |C'| - p|C| \ \big| \geq \varepsilon v p \ \right) = 
    \mathbb{P}\left( \ \big| \ |C'| - p|C| \ \big| \geq \delta \mu \ \right)\,$.
Now, recalling the definition~\eqref{eqn:lambda-def} of~$\lambda$, we verify the condition on the independence factor,~$\gamma$:
\begin{align*}
    \frac{c}{3} k \log m & = \frac{\varepsilon^2}{3} \lambda = \frac{\varepsilon}{3} \delta \mu 
    \leq e^{-1/3} \frac{\varepsilon}{2} \delta \mu && e^{-1/3} > 2/3\\
    & \leq e^{-1/3} \min(1, \delta) \delta \mu && \delta \geq \varepsilon/2 \text{ and } \varepsilon/2 \leq 1\,,
\end{align*}
where $|C| \leq \text{OPT} \leq 2v$ gives us the condition~$\delta \geq \varepsilon/2$. The condition $\varepsilon/2 \leq 1$ is arbitrary but recall that we want a $1-1/e-\varepsilon$ approximation so $\varepsilon < 1-1/e \leq 0.7$. Therefore, $\gamma \leq \lfloor \frac{c}{3} k \log m \rfloor \leq \lfloor e^{-1/3} \min(\delta, \delta^2) \mu \rfloor$ and Theorem~\ref{th:gen} gives the desired inequality.
\end{proof}

\section{Lemma \ref{lemma:nosk} (proof)}
\begin{lemma6}
Let $v$ be a guess in $\alg_\gamma$ and let $C'$ be the final subsampled coverage associated with guess $v$. If 
\begin{enumerate}
    \item $\big| \ |C'| - p|C| \ \big| < \varepsilon v p$; and
    \item $v \geq \text{OPT}$; and
    \item $(1-\varepsilon)(1-1/e-\varepsilon)\lambda \leq |C'|$;
\end{enumerate}
then $|C| > (1-1/e-\delta(\varepsilon))\text{OPT}$.
\end{lemma6}
\begin{proof}
Assuming condition~3, we have,
\begin{align*}
    |C'| - \varepsilon v p & \geq (1-\varepsilon)(1-1/e-\varepsilon) vp - \varepsilon v p\\
    |C|p & \geq (1-\varepsilon)(1-1/e-\varepsilon) vp - \varepsilon v p && \text{Condition 1}\\
    |C| & \geq (1-\varepsilon)(1-1/e-\varepsilon) v - \varepsilon v&& p > 0\\
    |C| & \geq (1-1/e-\delta(\varepsilon))v && \text{where } \delta(x)=x(3-1/e-x) \\
    |C| & \geq (1-1/e-\delta(\varepsilon))\,\text{OPT} && \text{Condition 2.}
\end{align*}
\end{proof}

\section{Proposition \ref{prop:nosk} (proof)}
\begin{proposition7}
    For $\gamma \geq \lfloor \tfrac{c}{3} k \log m \rfloor$, $\alg_\gamma'$ finds a $1-1/e-\delta(\varepsilon)$ approximation of the \textit{Maximum-$k$-Coverage} problem with probability at least $1 - 2 e/(k! m^{(c/6-1)k})$.
\end{proposition7}
\begin{proof}
    Let $v_s$ be the selected guess accordingly to procedure \ref{alg:findguess} and $v_*$ the right guess, i.e. $\text{OPT}/2 \leq v_* \leq \text{OPT}$. Also, we denote by $I_s$ the solution associated with guess $v_s$.
    \begin{itemize}
        \item If $v_s = v_*$, we already saw that the $1-1/e-\delta(\varepsilon)$ approximation is guaranteed if the event $E_{I_s} = \big| |C_s'| - p|C_s| \big| < \varepsilon v_s p$ is met.
        \item If $v_s > v_*$, then $v_s \geq \text{OPT}$ because each guess is of the form $2^g ||\mathcal{S}||_\infty$, so $v_s$ must be at least twice as big as $v_*$. Therefore, Lemma \ref{lemma:nosk} ensures the $1-1/e-\delta(\varepsilon)$ approximation if the event $E_{I_s}$ is met, because procedure \ref{alg:findguess} always takes a guess for which $|C_s'| \geq (1-\varepsilon)(1-1/e-\varepsilon)\lambda$.
    \end{itemize}
    To conclude, $\alg_\gamma'$ finds a $1-1/e-\delta(\varepsilon)$ approximation of \textit{Maximum-$k$-Coverage} if $v_s \geq v_*$ and $E_{I_s}$. Furthermore, a consequence of Corollary 9 in \S2.3 \cite{McGregor2018} is that, for the right guess $v_*$, if $\big| |C_*'| - p|C_*| \big| < \varepsilon v_* p$ then $|C_*'| \geq (1-\varepsilon)(1-1/e-\varepsilon)\lambda$, which makes the right guess a possible choice for procedure \ref{alg:findguess}. Therefore, $E_{I_*} \Rightarrow v_s \geq v_*$. Consequently:
    \begin{align*}
        \mathbb{P}\left(\{ v_s \geq v_* \} \cap E_{I_s}\right) & = 1 - \mathbb{P}\left(\{ v_s < v_* \} \cup \overline{E_{I_s}}\right) \geq 1 - \mathbb{P}\left(\overline{E_{I_*}} \cup \overline{E_{I_s}}\right) \\
        & \geq 1 - \mathbb{P}\left(\overline{E_{I_*}}\right) - \mathbb{P}\left(\overline{E_{I_s}}\right) \geq 1 - 2 \frac{e}{k! m^{(c/6-1)k}}
    \end{align*}    
\end{proof}

\section{Experimentation}

\begin{figure}[h]
    \centering
    \centerline{\includegraphics[width=\textwidth]{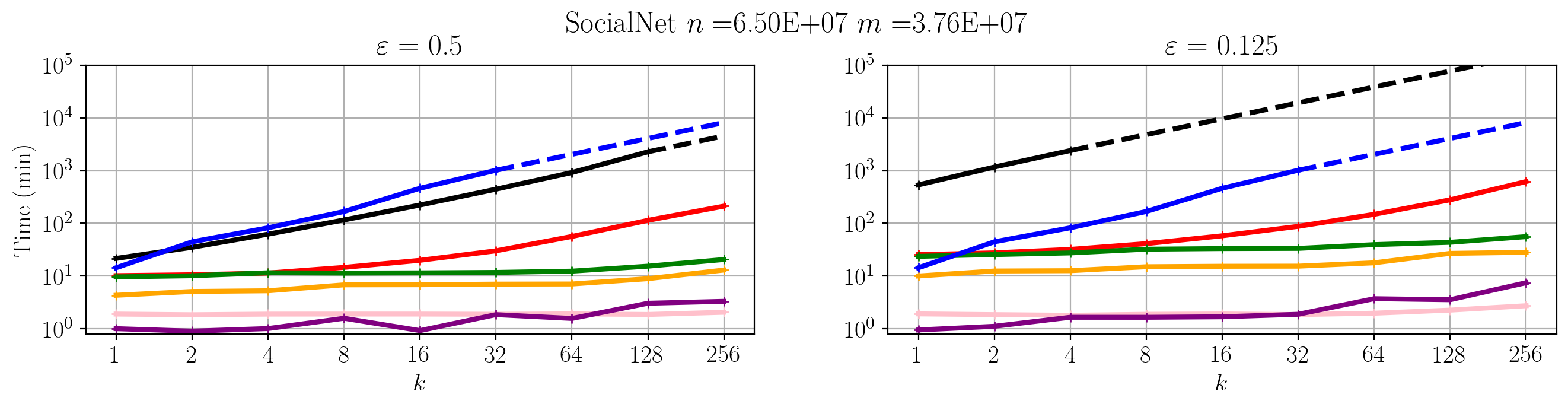}}
    \centerline{\includegraphics[width=\textwidth]{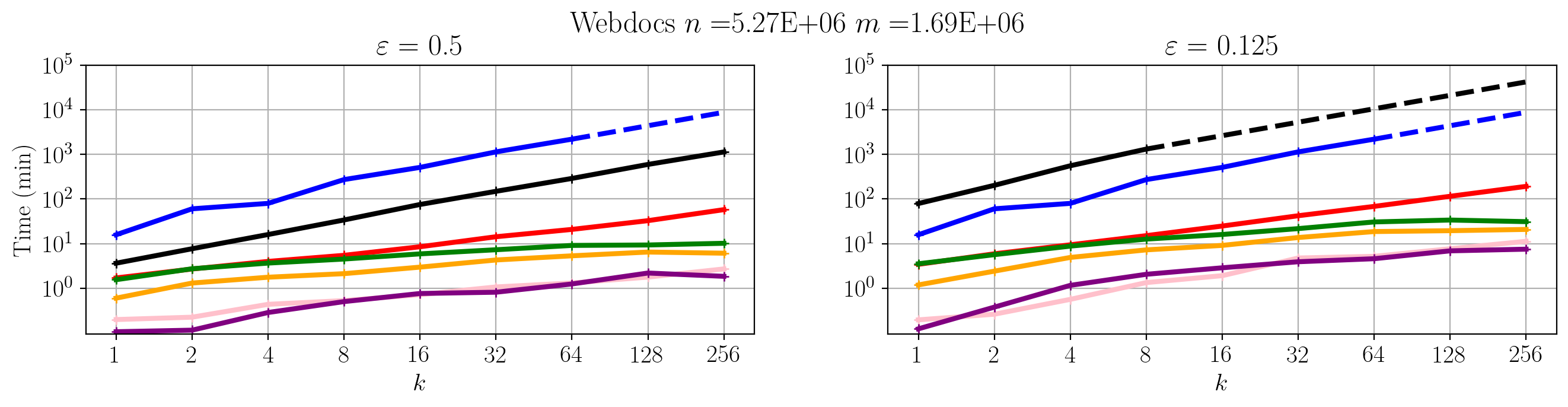}}
    \centerline{\includegraphics[width=\textwidth]{Figures/time_legend.png}}
    \caption{Running times of algorithms $\alg_\gamma'$, $\algfs'$, $\mathsf{SG}$, and $\mathsf{BMKK}$ on datasets \textit{SocialNet}, \textit{UKUnion} and \textit{Webdocs}.}
\end{figure}

\begin{figure}
    \centering
    \centerline{\includegraphics[width=\textwidth]{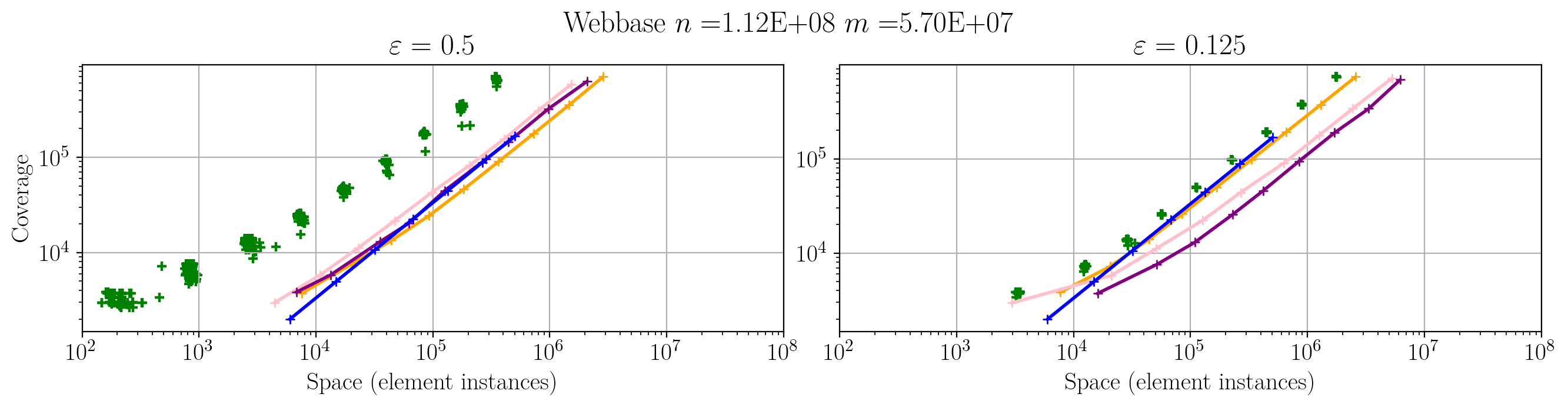}}
    \centerline{\includegraphics[width=\textwidth]{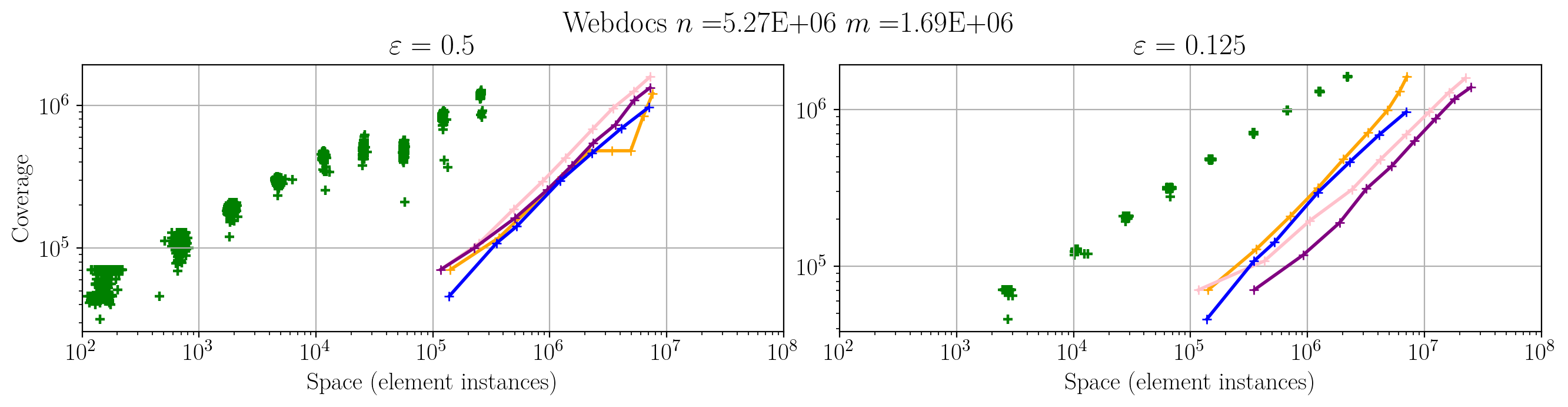}}
    \centerline{\includegraphics[width=\textwidth]{Figures/space_cov_legend.png}}
    \caption{Coverage versus space, for $k\in\{1, 2, 4, 8, 16, 32, 64, 128, 256\}$ on datasets \textit{Webbase} and \textit{Webdocs}. The anomalies of $\algfs'$ and $\alg_\gamma'$ in \textit{Webdocs} when $\varepsilon=0.5$ coincide with the coverage quality drop in Figure \ref{fig:cov}}
\end{figure}

\begin{figure}
    \centering
    \centerline{\includegraphics[width=\textwidth]{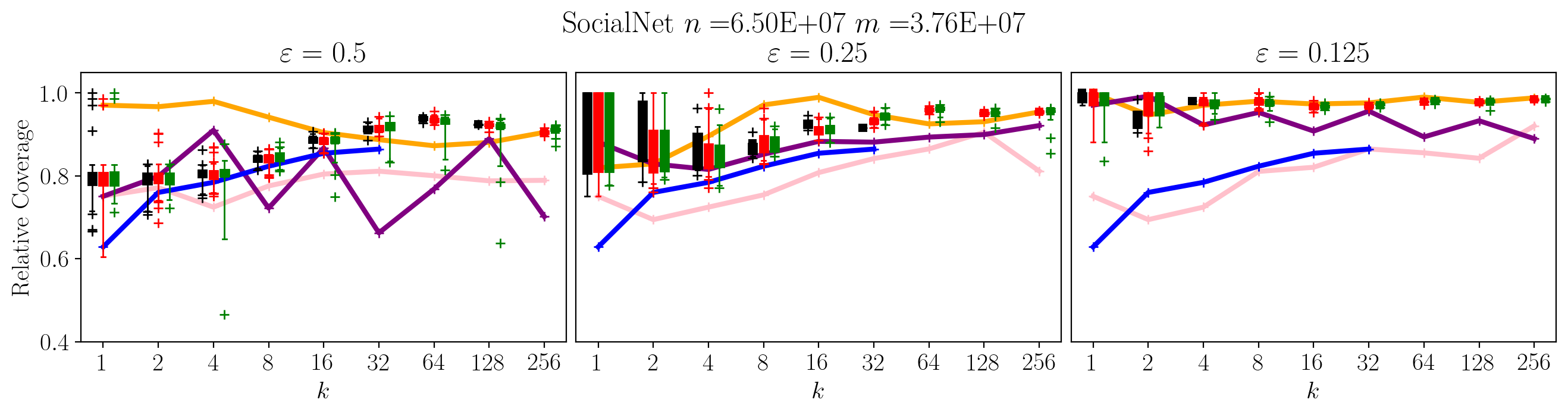}}

    \centerline{\includegraphics[width=\textwidth]{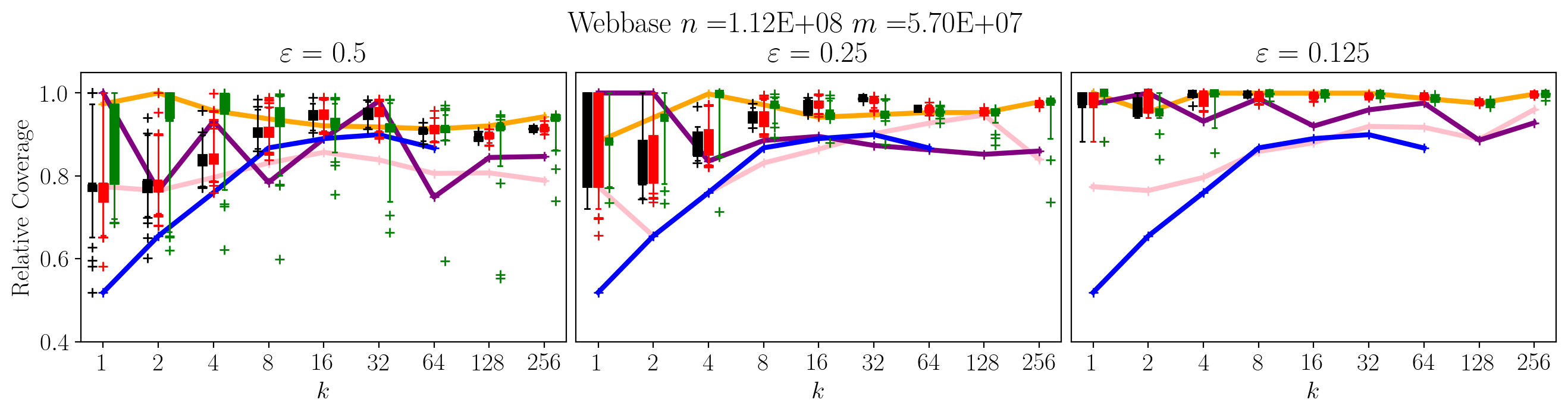}}

    \centerline{\includegraphics[width=\textwidth]{Figures/cov_legend.png}}
    \caption{Box plot of coverage produced by $\alg_\gamma'$, $\mathsf{SG}$ and $\mathsf{BMKK}$ relative to the coverage produced by the greedy algorithm for the datasets \textit{SocialNet} and \textit{Webbase}.}
\end{figure}

\begin{figure}
    \centering
    \centerline{\includegraphics[width=\textwidth]{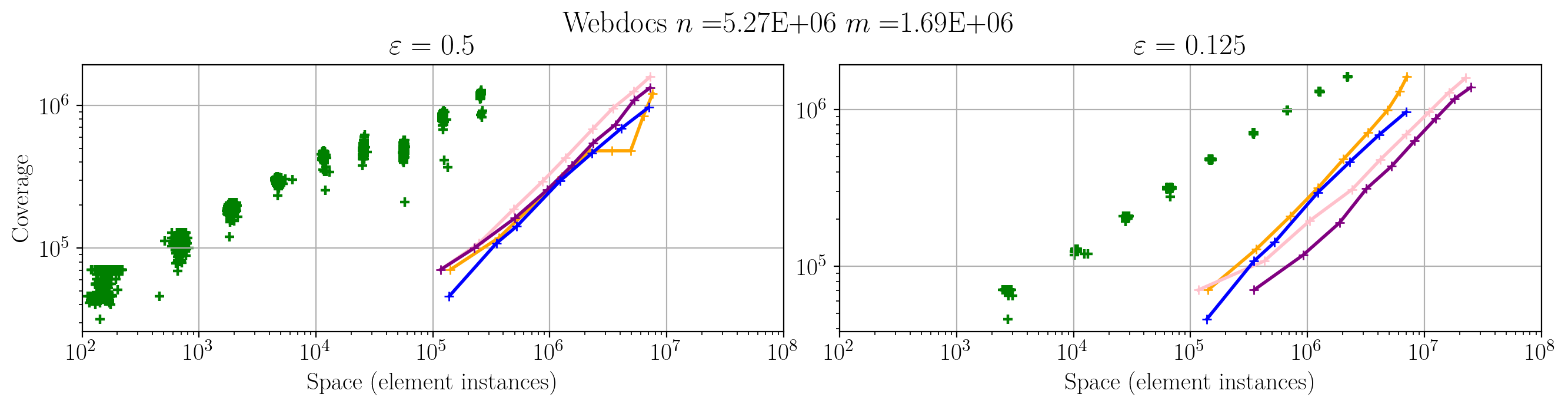}}
    \centerline{\includegraphics[width=\textwidth]{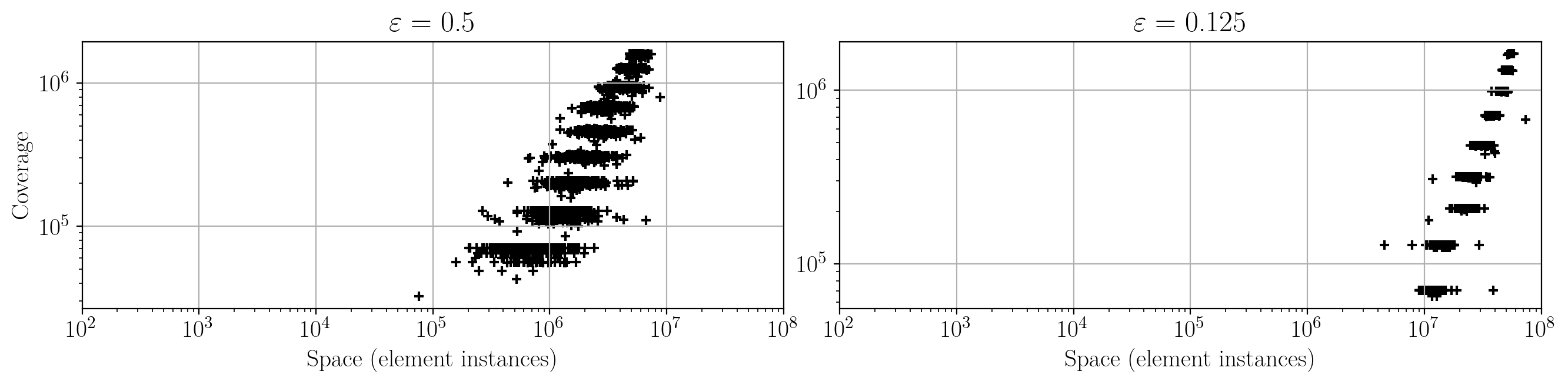}}
    \centerline{\includegraphics[width=\textwidth]{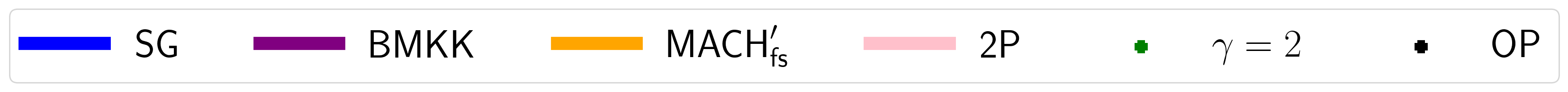}}
    \caption{Coverage versus space, the $\Tilde{\mathcal{O}}(\varepsilon^{-2} m)$ space algorithm $\mathsf{OP}$ \cite{McGregor2018} consumes systematically more space than all the other $\Tilde{\mathcal{O}}(\varepsilon^{-d} n)$ space alternatives. This is because the $\Tilde{\mathcal{O}}(\varepsilon^{-d} n)$ space algorithms actually scale linearly with respect to the returned coverage size with a hidden constant close to one. On the other hand, the $\Tilde{\mathcal{O}}(\varepsilon^{-d} m)$ space algorithms, such as $\mathsf{OP}$, have precisely a $\Tilde{\Theta}(\varepsilon^{-d} m)$ space complexity with a much bigger hidden constant, storing a fraction of each set.}
    \label{annex:cov-space-op}
\end{figure}

\end{document}